\documentclass[aps,prl,twocolumn,groupedaddress]{revtex4}

 \usepackage{graphicx}
 
\begin{document}


\title{Anisotropic flow  in 4.2A GeV/c C+Ta collisions}


\author{Lj. Simi\'c $^1$}
\author{J. Milo\v{s}evi\'c $^{1,2}$}
\email[]{simic@phy.bg.ac.yu}
\email[]{jmilos@physi.uni-heidelberg.de}
\affiliation{$^1$ Institute of Physics, P.O. Box 68, 11081 Belgrade, Yugoslavia\\
$^2$ Physikalisches Institut, Philosophenweg 12, 69115 Heidelberg, Germany}


\date{\today}

\begin{abstract}
 Anisotropic flow of protons and negative pions in 4.2A GeV/c C+Ta 
 collisions is studied  using the Fourier analysis of azimuthal 
 distributions. The protons exhibit pronounced directed flow. Directed 
 flow of pions is positive in the entire rapidity interval
 and indicates that the pions are preferentially emitted in the 
 reaction plane from the target to the projectile. The elliptic flow of 
 protons and negative pions is close to zero. Comparison with the 
 quark-gluon-string model (QGSM) and relativistic transport 
 model (ART 1.0) show that they both yield a flow signature similar to 
 the experimental data.
\end{abstract}


\maketitle

 The anisotropic transverse flow of particles has been actively studied 
 in nuclear collisions over a wide range of energies.
 At lower energies~\cite{DanO85,Part,Chance,Reis,Andro01}, 
   the flow is usually studied in terms of the  
 mean in-plane component of transverse momentum at a given rapidity, 
 $\langle p^x(y)\rangle$~\cite{DanO85}, and additionally quantified 
 in terms of derivative at  midrapidity $F_y=d\langle p^x \rangle/dy$.
 At high energies, the Fourier expansion of the 
 azimuthal distribution of particles constructed with respect to the reaction
 plane is used~\cite{Barrette94,VolZhang,PosVol}. In this expansion the 
 first harmonic, $v_1$, quantifies the directed flow while the second harmonic, 
 $v_2$, quantifies the elliptic flow. Using the Fourier expansion, the 
 anisotropic transverse flow was analyzed for heavy symmetric systems at the 
 AGS~\cite{Barrette94,Baret97a,Ajit98}, 
    SPS~\cite{Appel,Aggar}
 and RHIC~\cite{Acker} 
 energies. It was found that the flow observables are important 
 tools for 
 investigating properties of high density region created during the initial
 collisions. 
 In particular, the elliptic flow measurements may provide an important 
 constraint 
 on the Equation of State (EOS) of high density nuclear 
 matter~\cite{Pinken,Dan98}.

 In this paper the anisotropic transverse flow of protons and negative pions 
 in 4.2A GeV/c C+Ta collisions is studied using the Fourier analysis of
 azimuthal distributions. The analysis is performed using 1000 C+Ta 
 semicentral and central collisions obtained with the 2-m propane bubble 
 chamber, exposed at JINR, Dubna synchrophasotron. The semicentral and 
 central collisions are 
 selected by rejecting $\approx50\%$ events with the smallest multiplicity of 
 participant protons. 
  Additionally, the same type of analysis is performed using 100000 events 
 generated by the Quark-Gluon-String Model (QGSM)~\cite{Amel}
, and the same number of events generated by the 
 relativistic transport model (ART 1.0)~\cite{Bao}. 
 For these events the same centrality criterion is applied as in experiment, 
 leading to the average impact parameter $\langle b \rangle \approx$ 4.54 
  (4.05) fm 
 according to QGSM (ART 1.0). 
 
 In order to study the inelastic interactions with tantalum nucleus,
 ($^{181}$ Ta),
 three tantalum foils (1 mm thick and 93 mm apart) were 
 placed inside the chamber working in the 1.5-T magnetic field. The 
 characteristics of the chamber allow precise determination of the 
 multiplicity and momentum of all charged particles, as well as identification 
 of all negative and positive particles with momenta less than 0.5 GeV/c. All 
 recorded negative particles, except the identified electrons, are taken to be 
 $\pi^-$. Among them remains admixture of unidentified fast electrons 
 ($<$ 5\%) and negative strange particles ($<1\%$). All positive particles 
 with momenta less than 0.5 GeV/c are classified either as protons or $\pi^+$ 
 mesons according to their ionization density and range. Positive particles 
 above 0.5 GeV/c are taken to be protons, and because of this, the admixture 
 of $\pi^+$ of approximately 7\%  is subtracted statistically using the 
 number of $\pi^-$ mesons with $p>0.5$ GeV/c as follows: $n_p=n_{+}
 -n_{\pi^+}(p\leq 0.5\;\mathrm{GeV/c}) -0.82 \cdot
 n_{\pi^-}(p>0.5\;\mathrm{GeV/c})$, 
 where $n_+$ denotes the number of single positively charged particles, and 
 0.82 takes into account the proton deficit in tantalum nuclei and 
 consequently also $\pi^+$ deficit. From the ratio for each momentum interval 
 we determine the weight of protons which we further use when calculating 
 distributions of other kinematical variables. From the resulting number of 
 protons, the projectile spectators (protons with momenta $p>$ 3 GeV/c and 
 emission angle $\theta<4^{\rm o}$) and target spectators (protons with 
 momenta $p<$ 0.3 GeV/c) are further subtracted. The resulting number of 
 participant protons still contains some 17\%  of deuterons (with $p>$ 0.48 
 GeV/c) and $ 11\%$ of tritons (with $p>$ 0.65 GeV/c). The experimental data 
 are also corrected to the loss of particles emitted at small angles relative 
 to the optical axes of chamber and to the loss of particles absorbed by the 
 tantalum plates. The aim of this correction is to obtain isotropic 
 distribution in azimuthal angle and smooth distribution in emission angle 
 (both measured with respect to the direction of the incoming  projectile). 

 The azimuthal distribution of particles may be represented with the first 
 three terms of the corresponding Fourier expansion
 \begin{equation}
 \label{Fourexp}\frac{dN}{d\phi}\approx\frac1{2\pi}\big [1+2v_1\cos(\phi)+
 2v_2\cos(2\phi)\big ],
 \end{equation}
 where the two coefficients, $v_1$ and $v_2$, quantify the directed and 
 elliptic flow via $v_1=\langle cos(\phi) \rangle$ and 
 $v_2=\langle cos(2\phi) \rangle$. In Eq. (1), $\phi=\phi_{lab}-\Phi_{plane}$ 
 is the particle azimuthal angle determined with respect to the reaction 
 plane, with $\phi_{lab}$ denoting the azimuthal angle of particle in the 
 laboratory frame and $\Phi_{plane}$ denoting the azimuthal angle of the 
 (true) reaction plane. Since both the projectile momentum and the impact 
 parameter vectors are available in the QGSM/ART simulations, they are used to 
 determine the corresponding reaction plane. In the experiment the reaction
 plane is determined, for each event, using the projectile momentum vector and
 the vector $\bf Q$ determined from~\cite{DanO85} 
 \begin{equation}
 \label{vecQ} 
 {\bf Q}=\sum_i {\bf p}_{Ti} (y>y_{cm}+\delta)-\sum_j{\bf p}_{Tj}
 (y<y_{cm}-\delta), 
 \end{equation} 
 where ${\bf p}_{T}$ represents the transverse momentum of the proton emitted 
 in the forward ($y>y_{cm}+\delta$), or backward ($y<y_{cm}-\delta$), 
 hemisphere. Here, $y_{cm}$ denotes the center-of-mass rapidity of participant 
 protons while the quantity $\delta$ (=0.2) removes the protons emitted around 
 the $y_{cm}$ which are not contributing to the determination of the reaction 
 plane. The reaction plane angle for a proton is determined using this 
 expression only if this proton is not included in the above sum (i.e. if its 
 rapidity lies in the interval from $y_{cm}-\delta$ to $y_{cm}+\delta$). 
 Otherwise, in order to avoid autocorrelation (which is an effect of the 
 finite multiplicity), the $\bf Q$ vector is constructed by the analogous 
 expression in which the contribution of this proton is simply 
 omitted~\cite{DanO85}.
 We found that the reaction plane angle distribution is 
 essentially flat, thus confirming the absence of significant distortions 
 which could influence the magnitude of the extracted flow parameters. The 
 accuracy with which  the reaction plane angle is  determined, i.e. the 
 reaction plane resolution, is evaluated by the subevent 
 method~\cite{DanO85,PosVol}. 
 In this method, each event is divided randomly into two subevents, and then 
 the corresponding two reaction planes are determined. Subsequently, the 
 absolute value of the relative azimuthal angle, $\Phi_{12}$, between these 
 two estimated reaction planes is obtained. The relative azimuthal angle 
 distribution 
 is the basis for the correction of the Fourier coefficient, $v'_1$, obtained 
 with the estimated reaction plane. The relationship between the $v'_1$, and 
 the Fourier coefficient $v_1$ obtained relative to the true reaction plane, 
 is $v'_1=v_1\ \langle cos(\Delta\Phi)\rangle$, where  
 $\langle cos(\Delta\Phi)\rangle^{-1} $ is the correction factor 
 determined from 
 $\Phi_{12}$ distribution following the prescription 
 given in~\cite{PosVol,Ollit}. 
 \begin{figure}
  \includegraphics[width=8.57cm,height=8.57cm] {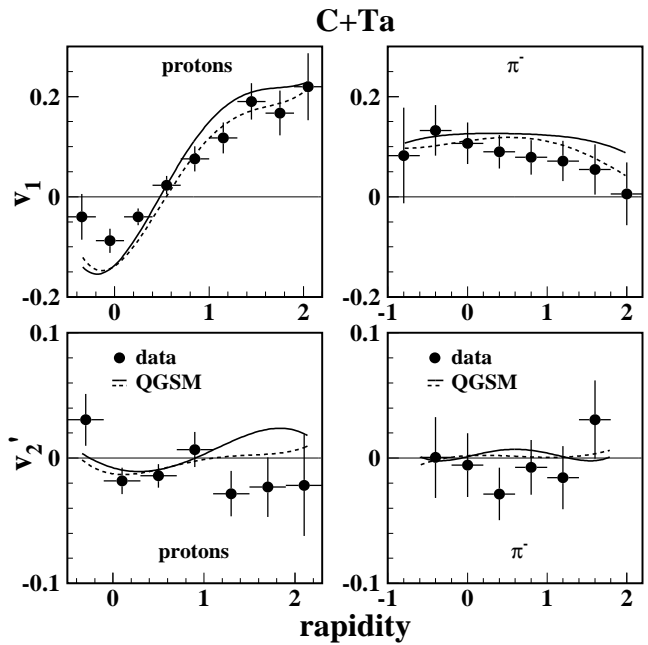}
  \caption{Rapidity dependence of $v_1$ and $v^{'}_2$ for protons and 
 $\pi^-$ for 4.2A GeV/c C+Ta collisions:
 $top$- filled circles represent the experimental results for $v_1$
 while the solid (dashed) line represents the    
 QGSM calculation for $v_1$ with respect to the true (estimated) 
 reaction plane;
 $bottom$- filled circles represent uncorrected experimental $v^{'}_2$ values 
 (see text), while the  
 solid (dashed) line represents the QGSM calculation for $v_2$ ($v^{'}_2$) 
 with respect to the true (estimated) reaction plane. \label{Fig. 1}}
  \end{figure} 
  We find $\langle cos(\Delta\Phi)\rangle$=0.59. 
 The correctness of this procedure is checked using the QGSM. Using this 
 model, the coefficient $v_1$ vs. rapidity, for protons and negative pions,
 is calculated with respect to 
 the true 
 reaction plane and also with respect to the estimated reaction plane. The 
 result of comparison is presented in Fig.1.
  
 The QGS calculations show 
 that the $v_1$ values 
 obtained with respect 
 to the estimated reaction plane, after applying correction procedure, are 
 somewhat underestimated around projectile rapidity for protons and 
  negative pions.  
  The QGS calculations  also show that the  correction procedure for 
  $v'_2$ as outlined above is not applicable because of the smallness of the 
  elliptic flow. Therefore, in the following analysis, this coefficient is not
  corrected to the reaction plane resolution.
   
  Fig. 1 (top) displays the experimentally determined $v_1$ coefficient vs. 
 $y$ (with $y$ calculated in the lab frame), 
 for protons and negative pions. In the case of protons the dependence 
 of $v_1$ on 
 rapidity is 
 characterized by a curve with a positive slope and with the zero-crossing at 
 $y \approx 0.5$, that corresponds to average rapidity of protons. The curve 
 indicates a 
 positive directed flow with magnitude 
 $v_1\approx 0.2$, at rapidities close to the  projectile rapidity 
  ($y_p=2.2$, at p=4.2A GeV/c).
 The
 QGSM reproduces satisfactorily the shape of $v_1(y)$ curve and within error 
 bars reproduces the magnitude of the flow. 
 The experimental 
 results are also compared with the relativistic transport model ART 1.0.
 These are shown in Fig. 2 where the calculations 
 are performed both for $\it stiff$ and $\it soft$ EOS. ART model yields a
 directed flow  which  flows trend similar to the 
 experimental data, but underestimates the flow intensity in the 
 projectile/target rapidity region. 
  
   \begin{figure}
 \includegraphics[width=8.57cm,height=8.57cm] {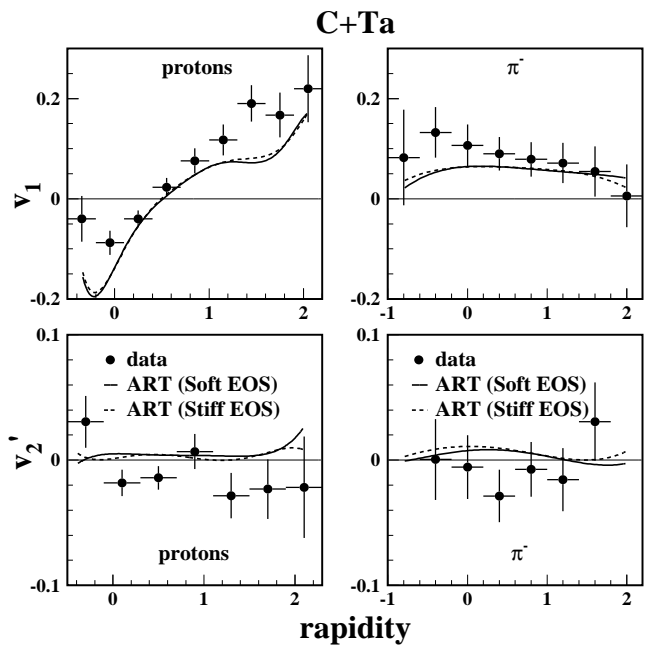}
  \caption{ Experimental results for $v_1$, $v^{'}_2$ as a function of 
  rapidity, compared with ART 1.0 model calculations.\label{Fig. 2}}
  \end{figure}

 Using the extracted values of 
 $v_1$ and their relation to the mean transverse momentum projected onto the 
 reaction plane, $v_1=\langle p_x \rangle /\langle p_T \rangle$, we can 
 evaluate $\langle p_x\rangle$ as a function of rapidity and determine the 
 slope, $F=d \langle p_x \rangle /d(y/y_b)$, with respect to rapidity 
 normalized to projectile rapidity in the lab frame. In the present 
 analysis we find for the 
 slope at mid-rapidity $F= 215\pm 32 $ MeV/c. Comparison with the other 
 results obtained at the same energy, for various C-nucleus combinations
 shows  increasing of the slope with the target mass: $F=$ 144 MeV/c for CC~\cite{Simic}, 134 MeV/c for CNe~\cite{Chaid} and 
 198 MeV/c for CCu~\cite{Chaid}.
 After the normalization 
 to the mass number of the colliding system we obtain the so-called scaled 
 flow $F_S = F/(A_1^{1/3} +A_2^{1/3})=27\pm 4$ MeV/c, 
 that allows a comparison of the energy dependence of flow values for different 
 projectile/target mass combinations.
 This value is in 
 agreement with the observed trend~\cite{Chance} 
 that after reaching the 
 maximum at a beam energy around 0.7-2A GeV, the directed flow slowly decreases 
 with increasing beam energy.

  For negative pions the experimental values of $v_1$ are positive in the 
  entire 
 rapidity interval. The $v_1$ is largest in the target rapidity region 
 ($v_1^{max}\approx0.10$) and monotonically decreases with increasing rapidity 
 towards the projectile rapidity. Such  
 $v_{1}$ dependence on $y$ reflects the fact that the pions are preferentially 
 emitted in the reaction plane from the target to the projectile. This 
 behavior is attributed to a shadowing effect  of the heavy target.
 Both QGSM and ART model cannot strictly account for the $v_{1}(y)$ 
 dependence for negative pions. 
 
 Figure 1 (bottom) displays the experimentally determined $v'_2$ 
 coefficient versus $y$ for protons and negative pions. The uncorrected 
 values of $v'_2$ show that in the entire 
 rapidity interval the elliptic flow is small $|v'_{2}| \leq 0.02$ if not zero, 
 and this is consistent 
 with  QGSM (Fig. 1) and ART (Fig. 2) predictions.
 The  values of $v'_2$ 
 additionally confirm the result~\cite{Pinken}  that at beam energy 
 of $\approx 4$ GeV the 
 elliptic flow exhibits a transition from negative (out-of-plane) to positive
 (in-plane) elliptic flow. 
 
   \begin{figure}
 \includegraphics[width=8.57cm,height=5cm] {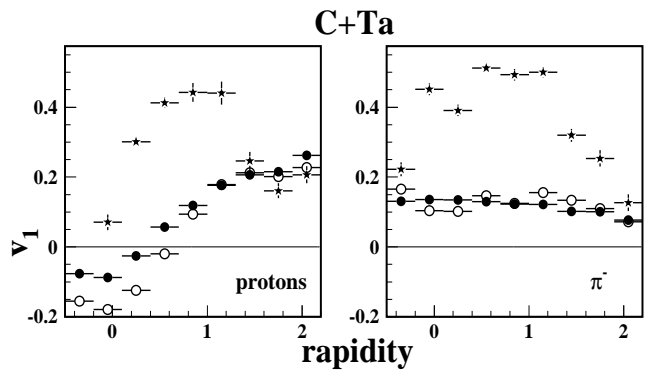}
  \caption{Rapidity dependence of $v_1$ for protons and 
  $\pi^-$  originating from:
   primary non-resonant interactions ({\it stars}),
   decay of resonances  ({\it full circles}), and secondary
   non-resonant interactions ({\it open circles}),
   for 4.2A GeV/c C+Ta collisions generated 
   with the QGSM. \label{Fig. 3}}
  \end{figure}
  
 Since the QGSM predictions are in fair agreement with the various experimental 
 results at 4.2A GeV/c, we use this model to clarify the question of which of 
 the  processes are responsible for the flow effect. In this model, for C+Ta 
 collisions $\approx 43\%$ of protons and $\approx 83\%$ of $\pi^-$  originate 
 from decay of the lowest-lying resonances ($\Delta'$s, $\varrho,\omega,\eta$ 
 and $\eta^{'}$). The rest originates from the 'non-resonant' primary and 
 secondary interactions of the type: $NN\rightarrow NN\pi$, $\Delta
 N\rightarrow \Delta N$, $\pi N \rightarrow \pi N$, $\pi NN\rightarrow NN$. The
 protons and pions from primary interaction escape the collision zone without 
 further rescattering and comprise $\approx 1 \%$ of the total.  
 According to QGSM, we separately evaluate the flow of protons and pions 
 originating from the following sources:\\
 ({\it i}) decay of resonances;\\
 ({\it  ii }) primary non-resonant interactions;\\
 ({\it iii }) and secondary non-resonant interactions.\\

 Figure. 3 shows  $v_1$ vs. rapidity for protons and negative pions 
 originating from the decay of resonances, and from primary and secondary 
 non-resonant interactions. The protons originating both from the decay of 
 resonances and from the secondary non-resonant interactions show a 
 directed flow of similar intensity. The same applies to the flow of pions. 
 For both protons and pions from the primary interactions  the directed flow 
 has a maximum around $y\approx 0.6$, and decreases towards 
 the projectile and target 
 rapidities. These protons and pions are produced at the early stage of 
 the collision, and both are shadowed by the cold spectators. 
  Later, after the spectator matter leaves the collision zone, rescattering of
  protons near the beam (target) rapidity region is small, 
 while the pions are still affected with the shadowing effect of the 
 participant nucleons trough both pion rescattering and reabsorptions.
  This could be the underlying mechanism  that  leads to the different 
  intensity and dependence on rapidity for the directed flow of 
  pions and protons.

 In summary, the directed and elliptic flow of protons and negative pions in 
 4.2A GeV/c C+Ta collisions was examined using the Fourier analysis of 
 azimuthal distributions of experimental events and also by using the events 
 generated  
 by the QGSM and ART 1.0. The protons exhibit strong directed 
 flow with 
 magnitude  $v_1\approx 0.2$ at rapidities close to the projectile rapidity.
 The directed flow of pions is positive in the entire rapidity and slightly
 peaked at target rapidity, where $v_1\approx 0.1$. This behavior
 indicates that the pions are preferentially emitted in the 
 reaction plane from the target to the projectile.
 For both sets of particles in the entire rapidity interval the elliptic 
 flow is 
 close to zero ($|v^{'}_{2}|\leq 0.02$), this being consistent with the result 
 that at beam energy 
 of $\approx 4$ GeV the 
 elliptic flow shows a transition from negative  to positive.
  A comparison with the 
 quark-gluon-string model (QGSM) and relativistic transport 
  model (ART 1.0) shows that they both yields the flow signature similar to 
 the experimental data. Additionally, the QGSM shows that  two factors 
 that dominantly 
 determine the proton and pion flow at this energy, are the decay of 
 resonances and the rescattering of secondaries. The shadowing by the cold 
 spectator matter affects only the flow of the particles produced at the early 
 stage of the collision.\\

\begin{acknowledgments}
The authors are grateful to members of the JINR Dubna group that 
participated in data processing.  
\end{acknowledgments}

\end{document}